\documentclass[acmsmall]{acmart}
\AtBeginDocument{%
  \providecommand\BibTeX{{%
    \normalfont B\kern-0.5em{\scshape i\kern-0.25em b}\kern-0.8em\TeX}}}

\setcopyright{acmcopyright}
\acmJournal{PACMHCI}
\acmYear{2022} \acmVolume{6} \acmNumber{CSCW2} \acmArticle{481} \acmMonth{11} \acmPrice{15.00}\acmDOI{10.1145/3555582}

\usepackage{comment}
\usepackage{caption}
\usepackage{subcaption}
\usepackage{fontawesome}
\usepackage{caption}
\usepackage{dblfloatfix}
\usepackage[normalem]{ulem}

\newcommand{\change}[1]{{\color{black}#1}}
\newcommand{\minorchange}[1]{{\color{black}#1}}

\begin{document}

\title{“I was Confused by It; It was Confused by Me:” Exploring the Experiences of People with Visual Impairments around Mobile Service Robots}

\renewcommand{\shorttitle}{Experiences of People with Visual Impairments around Mobile Service Robots}
\author{Prajna M. Bhat}
\email{pmbhat@wisc.edu}
\affiliation{%
 \institution{University of Wisconsin-Madison}
 \city{Madison}
\state{Wisconsin}
 \country{USA}
}

\author{Yuhang Zhao}
\affiliation{%
    \institution{University of Wisconsin-Madison}
    \city{Madison}
    \state{Wisconsin}
    \country{USA}
}
\email{yuhang.zhao@cs.wisc.edu}


\begin{abstract}
  Mobile service robots have become increasingly ubiquitous. However, these robots can pose potential accessibility issues and safety concerns to people with visual impairments (PVI). We sought to explore the challenges faced by PVI around mainstream mobile service robots and identify their needs. Seventeen PVI were interviewed about their experiences with three emerging robots: vacuum robots, delivery robots, and drones. \change{We comprehensively investigated PVI's robot experiences by considering their different roles around robots}---direct users and bystanders. Our study highlighted participants' challenges and concerns about the accessibility, safety, and privacy issues around mobile service robots. We found that the lack of accessible feedback made it difficult for PVI to precisely control, locate, and track the status of the robots. Moreover, encountering mobile robots as bystanders confused and even scared the participants, presenting safety and privacy barriers. We further distilled design considerations for more accessible and safe robots for PVI.
\end{abstract}

\ccsdesc[500]{Human-centered computing~Accessibility}
\ccsdesc[500]{Computer systems organization~Robotics}

\keywords{Accessibility, visual impairments, mobile service robots, human-robot interaction}


\maketitle

\section{Introduction}

Service robots refer to robots that ``perform useful tasks for humans or equipment excluding industrial automation applications'' \cite{ServiceRobotDef}. Both personal (e.g., vacuum robots) and professional service robots (e.g., delivery robots) are becoming prevalent and have been adopted at various places such as homes, schools, and hospitals. They assist humans by automatically completing dirty, distant, dangerous, and repetitive tasks \cite{ServiceRobotDefn}, offering lower costs and risks. To perform these tasks, most service robots are mobile, which can move around autonomously or by following users' control.  

The market of service robots is growing rapidly. The unit sales of service robots for personal and domestic use increased by 34\% from 17.8 million in 2018 to more than 23.2 million in 2019 \cite{ServiceRobotRecord}. Moreover, the need for contactless services in the COVID-19 pandemic has increased the adoption of professional service robots. For example, hospitals in the United States have deployed service robots (e.g., Moxi) to reduce the stress and exposure risk for healthcare workers \cite{HightechHelper}. Domino's has also started providing contact-free pizza delivery service via R2 robots \cite{DominoDelivery}. Prior research has shown that people are more willing to visit a hospital or a restaurant with service robots during the pandemic \cite{wan2020robots}. Following this trend, it is projected that the market size of service robots will increase to \$102.5 billion in 2025 \cite{RobotTrend-Covid19}.

While presenting great potential, mobile service robots can pose various challenges to people with visual impairments (PVI). Most mainstream robots focus on visual aesthetics and provide only visual interfaces, bringing accessibility barriers to PVI.
For example, a recent study by Gadiraju et al. showed that PVI struggled with controlling their drones due to the lack of accessible feedback of the drone's position and behaviors \cite{gadiraju2021fascinating}. Beyond accessibility issues, mobile service robots can also raise safety concerns. For example, a vacuum robot at home could become a low-lying obstacle for PVI during indoor navigation, causing tripping injuries. 

Besides interacting with the robots as direct users, PVI may also come across other people's robots or public-use service robots. As a bystander (\change{i.e., a non-user who can be affected by the technology, but do not have control over it \cite{bernd2020bystanders}}), PVI face more severe challenges and risks due to the lack of control and understanding of the robots. For example, a delivery robot may occupy the sidewalk and become a safety hazard for pedestrians with disabilities \cite{bennett2021accessibility}. Moreover, the prevalence of sensors (e.g., camera, GPS) on robots can pose severe privacy threats to bystanders with visual impairments since they cannot visually detect the robots and avoid being captured.
All these potential issues can prevent PVI from accepting, adopting, and protecting themselves from the mainstream service robots. With the increasing adoption of mobile service robots, a timely study is needed to thoroughly investigate PVI’s challenges and needs around different mainstream mobile service robots, thus providing actionable design guidelines to foster the accessibility of this emerging technology.

To fulfil this need, we conducted a semi-structured interview study with 17 PVI to develop a deep understanding of their interactions with different mobile service robots. Unlike most prior work that focused on the direct use of robots \cite{gadiraju2021fascinating,azenkot2016enabling}, we comprehensively considered PVI's different possible roles around robots---\textit{\textbf{direct users}} and \textit{\textbf{bystanders}}---to fully understand the interaction dynamics between PVI and robots. \change{Specifically, we aim to answer two research questions:

\textbf{RQ1}: What are PVI's experiences, challenges, and needs when they directly use different mobile service robots, including both personal robots and shared robots?

\textbf{RQ2}: What are PVI's experiences, challenges, and needs when they encounter mobile service robots as bystanders?}

Our study focused on three representative mobile service robots: vacuum robots, delivery robots, and drones, which are either commonly-used or emerging robots. For each type of robots, we interviewed participants \minorchange{about their} experiences from both direct users' and bystanders' perspectives. 
\change{Our study developed a comprehensive and thorough understanding of PVI's robot experiences. As direct users, participants faced various accessibility issues across the whole human-robot interaction process, from initiating, to tracking, controlling, to eventually stopping the robots. 
Fine-grained control of the robots was particularly challenging due to the lack of precise feedback. As bystanders around robots, participants expressed safety and privacy concerns, such as themselves accidentally tripping over a low-lying robot, a moving robot breaking their cane or confusing the guide dog, and their personal space being invaded by other people's robots. 
}

In summary, we contribute to the CSCW community by presenting an in-depth exploration of PVI's experiences with different mobile service robots. We investigated PVI-robot interaction through the lens of direct users and bystanders, revealing the accessibility, safety, and privacy challenges faced by PVI as different stakeholders around robots. We further derived a set of design considerations to inspire more accessible and safe robot design for PVI.

\section{Related Work}

\subsection{Sighted People's Experiences and Perceptions of Mobile Service Robots}

Service robots perform challenging and repetitive tasks for humans \cite{ServiceRobotDef}. Such robots range from partial autonomy that involves human-robot interaction to full autonomy without active human intervention. Service robots can be classified into two major categories according to its use and operators---personal service robots and professional service robots \cite{zielinska2016professional}. Personal service robots are used for non-commercial tasks by lay persons, such as domestic vacuum robots, remote presence robots, and automated wheelchairs, while professional service robots are usually operated by properly trained experts for commercial tasks, such as delivery robots and surgery robots. 

A wealth of prior research has explored people’s experiences and attitudes when interacting with different service robots \cite{young2009toward,schneiders2021domestic, dautenhahn2005robot}, such as domestic vacuum robots \cite{forlizzi2006service, de2019would, sung2007my}, delivery robots \cite{jennings2019study, lee2020robots, pani2020evaluating}, and drones \cite{culver2014battlefield, eissfeldt2020acceptance,lidynia2017droning}. For example, Forlizzi and Disalvo \cite{forlizzi2006service} conducted ethnographic studies with 14 families to explore the effects of a Roomba vacuum robot on home ecology. They found that the adoption of Roomba changed people's cleaning practices by enabling multitasking and planned cleaning, as well as influencing the roles of family members in housekeeping by engaging men and children in cleaning. Delivery robot is an emerging shared robot that has recently attracted HCI researchers' attention, especially during the COVID-19 pandemic. Via a survey with 473 participants, Pani et al. \cite{pani2020evaluating} explored the consumers’ preferences, trust, and attitudes towards the deployment of autonomous delivery robots during the pandemic. The study showed an overall positive attitude towards the delivery robots with 61.28\% customers indicating willingness to pay for the robot service. Moreover, Lee and Toombs \cite{lee2020robots} explored the public perceptions of Starship delivery robots by analyzing Reddit posts, finding that people perceived the robots differently as mascots, friends, or aggressive gangs, and expressed affections for the robots.

Besides the direct interaction with robots as users, researchers have also explored people's experiences and attitudes when they come across publicly used service robots as bystanders, revealing their concerns around the robots. For example, Salvini et al. \cite{salvini2021safety} pointed out that pedestrians faced both physical and psychological safety hazards in the presence of mobile robots in public space due to the lack of knowledge of properly interacting with the robots and the lack of clarity of the robots' intentions. They highlighted the needs for standards and guidelines for public-use robots that protect bystanders' safety and rights. Thomasen \cite{thomasen2020robots} also argued that civilians' space and comfort must be prioritized over robots' use of public space. In addition to safety concerns, researchers have also analyzed the potential privacy risks posed by the publicly used robots \cite{butt2021deployment, LINTAN2021101462}. For example, Tan et al. \cite{LINTAN2021101462} studied 1050 Singaporean residents' responses to a questionnaire, asking about their experiences of encountering flying drones in public space. They found that people's primary fears were unauthorized filming by the drones and loss of privacy. 

While prior research has thoroughly investigated people's robot experiences and concerns from both direct user's and bystander's perspectives, this research has mainly focused on people who are sighted. Unlike sighted people, PVI can face additional accessibility challenges and more severe safety and privacy risks when using and encountering mobile service robots. However, their unique challenges and needs remain understudied.

\subsection{Specialized Assistive Service Robots for People with Visual Impairments}

Robotics research for PVI has mainly focused on designing and building specialized assistive robots to facilitate various daily tasks. Most assistive robots were designed to enable PVI to navigate surrounding environments \cite{guerreiro2019cabot, kulyukin2005robocart, tachi1984guide, kulyukin2006robot}. For example, Guerreiro et al. built CaBot \cite{guerreiro2019cabot}, a suitcase shaped robot that \minorchange{guided} blind people via vibro-tactile feedback to help them avoid obstacles on the path and arrive \minorchange{at} a destination safely. They evaluated CaBot with ten blind participants who performed navigation tasks, finding that users felt comfortable, safe, and confident when navigating with CaBot. Similarly, a suitcase system BBeep \cite{kayukawa2019bbeep} was designed to assist blind pedestrians with collision avoidance via audio feedback. Moreover, some guide robots were designed for specific scenarios, such as a grocery store. For example, RoboCart \cite{kulyukin2005robocart} was a robotic shopping assistant designed to help PVI navigate a grocery store while carrying purchased items. Besides on-the-ground robots, researchers also leveraged flying drones to guide visually impaired users during navigation \cite{avila2015dronenavigator, al2016exploring}. For example, Avila et al. \cite{avila2015dronenavigator} proposed a wearable bracelet prototype with a small drone mounted on it. When a blind user verbally assigned a navigation task to the drone, the drone started flying and guided the user to the destination. 

In addition to navigation, researchers have also designed robotics technology to support PVI in other tasks. For example, Huppert et al. \cite{huppert2021guidecopter} designed GuideCopter, a drone-based system to assist PVI in an object localization task. With GuideCopter, a drone was tethered to a blind user's finger, thus providing fine-grained haptic feedback to guide the user's hand to a target object. Moreover, Bonani et al. \cite{bonani2018my} explored the potential of a collaborative assistive robot that provided physical instructions and feedback to guide blind users' hands in an assembly task. In an evaluation with 12 visually impaired participants, they found that participants perceived the physical assistive robots to be warmer, more competent, and more useful than a voice-only assistant.  

While various specialized assistive robots have been designed to empower PVI, the mainstream service robots, which are more widely used in all kinds of environments and tasks, can still pose severe barriers to PVI, further marginalizing them as the robotics technology advances.

\subsection{People with Visual Impairments and Mainstream Service Robots}

With the growing popularity of service robots, some recent research has started exploring PVI's expectations and needs for the mainstream robots. For example, to enable building service robots to guide blind people properly, Azenkot et al. \cite{azenkot2016enabling} conducted participatory design sessions, where three designers and five visually impaired users collaboratively designed the interaction and behaviors for a guide robot via brainstorming and discussions. The study specified design recommendations for different PVI-robot interaction stages, such as how to initiate contact with a visually impaired user, how to properly guide the user, and how to alert them of obstacles and turns. Instead of focusing on one specific type of robot, Bonani et al. \cite{bonani2018my} explored blind people's perceptions and visions on general service robots via a set of focus group discussions. They found that while having concerns about robot control, reliability, and maintenance, participants expected that robots would be integrated in their daily life and assist with various tasks such as navigation, housekeeping, and social interactions. However, these studies focused on discussing the ideal service robots for visually impaired users without investigating their real-world experiences with robots. In fact, Bonani et al. clarified that most of their participants had never interacted with a real robot \cite{bonani2018my}. 

Moreover, Qbilat et al. \cite{qbilat2021proposal} proposed a set of accessibility guidelines for human-robot interaction by adapting and extending existing accessibility standards and guidelines for other mainstream user interfaces (e.g., web sites, mobile applications). The guidelines were evaluated by seventeen human-robot interaction (HRI) designers from four aspects, including usability, social acceptance, user experience, and social impact. Most designers found the guidelines helpful and were willing to use them in their future design. Nonetheless, no people with disabilities were involved in the development and evaluation of the guidelines. 

There has been little work exploring PVI's firsthand experiences and challenges around the mainstream service robots. Gadiraju et al. \cite{gadiraju2021fascinating} surveyed 59 PVI and interviewed 13 of them to understand their motivation and experiences with drone piloting. They found that PVI showed a general curiosity about the drones and flew drones for different reasons, such as engaging in family activities and collecting environment information. The study also highlighted participants' challenges and preferences for different drone control approaches, suggesting that multiple control and feedback configurations should be supported based on users' environments and preferences. Further, Bennett et al. \cite{bennett2021accessibility} touched on the safety concerns of pedestrians with motor-related disabilities when encountering delivery robots on the sidewalk. Via interviews with activists with disabilities, government officials, and commercial representatives, they revealed the barriers posed by the evolving micromobility technology (e.g., bikeshare) and the related policies for people with disabilities. However, their work mainly focused on the policy and regulation perspective instead of the actual experience with service robots. Only one participant who was a power wheelchair user shared her experience with delivery robots. Thus, there is a lack of knowledge in the accessibility field on how PVI negotiate the growing number of service robots in their daily life. 

Our research sought to fill the gap in prior research by deeply investigating PVI's experiences and needs around different mainstream service robots, from both direct users' and bystanders' point of view, thus deriving design guidelines for more accessible and safe human-robot interaction.

\section{Method}
\subsection{Participants}
We recruited 17 participants with visual impairments (11 female, 6 male), whose ages ranged from 24 to 79 ($mean = 49$, $SD = 15.78$). All participants were legally blind\footnote{Legally blind means that a person's best-corrected visual acuity in her better eye is 20/200 or worse, or her visual field is 20 degrees or narrower. 
}: 13 were blind (i.e., little to \minorchange{no} vision), and 4 had low vision (Table \ref{tab:demo}). 

Participants were recruited via the National Federation of the Blind (NFB). After being reviewed by the NFB research advisory council, our recruitment information was posted to the NFB member list. We attached a screening survey link to our recruitment email, asking about potential participants’ age, visual abilities, and their experiences with service robots (Appendix \ref{appendix: screen}). Participants were eligible for our study if they were 18+ years old, had visual impairments, and had experience with at least one type of service robots among vacuum robots, delivery robots, and drones. 

\begin{table}
    \scriptsize
  \caption{Participants' demographics \change{and robot experiences} (\faUser - \change{Direct} User, \faStreetView - Bystander).}
  \label{tab:demo}
  \begin{tabular}{p{0.7cm} p{0.9cm} p{1.2cm} p{3.4cm} p{3.1cm} p{2.8cm}}
    \toprule 
    \textbf{Pseu-donym} & \textbf{Age/ Gender} & \textbf{Visual \newline Ability} & \textbf{Experience with \newline Vacuum Robots} & \textbf{Experience with 
    \newline Delivery Robots} & \textbf{Experience with \newline Drones}\\
    
    \toprule 
    
    Mina & 29/F & Blind & \faUser: \change{twice a week for 5 months} & - & - \\
    \midrule
    
    Meg & 38/F & Blind & \faUser : \change{daily use for 2 yrs}  & -  & \faUser : \change{once indoors}\\
     & & & \faStreetView : \change{once at a friend's place} & &\\
    \midrule
    
    Edward & 42/M &  Blind & \faUser : \change{several times a week for 2 yrs} & - & - \\
    \midrule
    
    Summer & 39/F & Low Vision & \faUser : \change{once a week for 2 yrs} & \faStreetView : \change{once at college campus}  & \faStreetView : \change{once outdoors}\\
    \midrule
    
    Jen & 49/F & Blind & \faUser : \change{alternate days for 1 yr} & - & - \\
    \midrule
    
    Ara & 38/F & Low vision & \faUser : \change{4 times since a few months ago} & \faUser : \change{intermittently for 2 yrs}  & \faStreetView : \change{a couple of times per yr}\\
     & & & & \faStreetView : \change{everyday at college campus} & \\
     \midrule
    
    Sara & 62/F & Blind & \faUser : \change{daily use for 3 yrs} & - & \faUser : \change{once with grandchildren}  \\
     \midrule
    
    Rita & 51/F & Blind & \faUser : \change{tried for 2 weeks} & - & \faUser : \change{once in a park} \\
     & & & & & \faStreetView : \change{more than once}\\
     \midrule
    
    Kat & 44/F & Blind & \faUser : \change{daily use for 3-4 yrs} & \faStreetView : \change{regularly on sidewalks} & -\\
    \midrule
    
    Thomas & 79/M & Blind & \faUser : \change{twice since 2 yrs ago} & \faStreetView : \change{three times on sidewalk} & \faStreetView : \change{several times in public}\\
    \midrule
    
    Ahmed & 36/M & Blind & \faUser : \change{once 2 yrs ago} & -
    & \faUser : \change{once in a park} \\
     & & & & & \faStreetView : \change{once at a wedding}\\
     \midrule
    
    Rose & 24/F & Blind & \faUser : \change{alternate days for 2-3 yrs} & \faStreetView : \change{often on sidewalks}  & \faStreetView : \change{5 times in public}\\
    \midrule
    
    Samuel & 54/M & Low vision & \faUser : \change{daily use for 6 months} & - & \faUser : \change{once indoors and \newline once outdoors}\\
     & & & & & \faStreetView : \change{once or twice per year}\\

    \midrule
    
    Tim & 39/M & Blind & \faUser : \change{daily use for a yr}  & - & - \\
    \midrule
    
    Reena & 27/F & Blind & \faUser : \change{several times a week for 3 yrs}  & - & \faUser : \change{once indoors}\\
     & & & \faStreetView : \change{a few times at friends' place} & &\\
     \midrule
    
    John & 35/M & Low vision & \faUser : 
    \change{purchased in 2018;} & \faUser : \change{4 times since a yr ago} & \faStreetView : \change{once in a park}\\
     & & & seldom use since 2020 & \faStreetView : \change{several times on sidewalk} & \\
     \midrule
    
    Nina & 37/F & Blind & \faUser : \change{1-2 times per week for 1 yr}  & -  & \faUser : \change{once in the backyard} \\
     & & & \faStreetView: \change{once at a friend's place} &  & \faStreetView: \change{once in a park}  \\
    
  \bottomrule
\end{tabular}
\end{table}


We received 54 survey responses. To collect sufficient data for each type of robots, we prioritized respondents who had experienced (i.e., used or encountered) multiple types of robots during our recruitment. We first contacted all respondents with experiences of more than one type of robots, and thirteen of them participated in our study. We then contacted the rest of the participants (they only had experience with vacuum robots). Another four participants \minorchange{responded us within our one-month recruitment window (April 22nd - May 30th, 2021) and participated in} the study. 

\begin{figure}[h]
    \centering
    \includegraphics[width=0.85\linewidth]{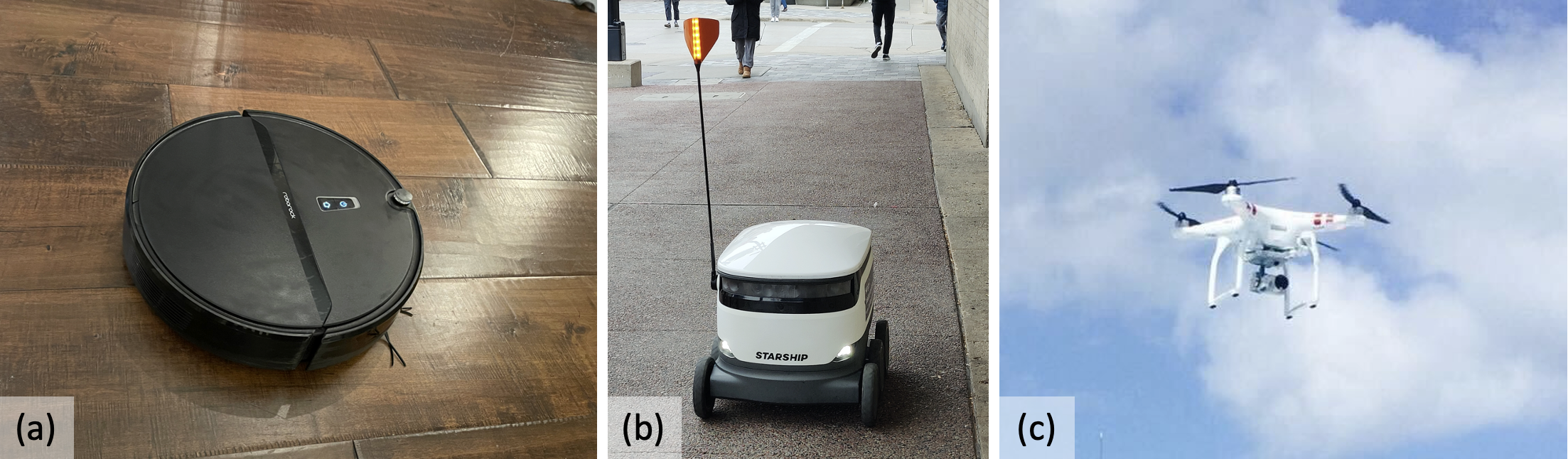}
    \caption{Three mobile service robots we focused on: (a) a vacuum robot; (b) a delivery robot; (c) a flying drone.}
    \label{fig:threerobots}
    \vspace{-13pt}
\end{figure}

\subsection{Three Types of Mobile Service Robots}
We focused on three types of service robots: vacuum robots, delivery robots, and drones (Fig. \ref{fig:threerobots}). Our robot selection followed three criteria: (1) the robots should be either commonly-used (e.g., vacuum robots \cite{VacuumRobotStats}) or emerging (e.g., delivery robots \cite{DeliveryRobotStats} and drones \cite{DroneStats}) to represent the mainstream robots; (2) the robots should be mobile robots that can move or fly around since they can pose more safety and privacy concerns for PVI than stationary robots; and (3) the robot selection should have a good coverage---we covered different robot use (vacuum robots and drones for personal use, and delivery robots for professional and shared use) and mobility styles (vacuum robots and delivery robots are low-lying mobile robots, while drones are flying robots), thus comprehensively reflecting PVI' experiences and needs around various service robots. We introduce the definition and the basic features of these robots:


\begin{itemize}
    \item {\textbf{Vacuum robots} refer to autonomous robotic vacuum cleaners that move around and sweep up dirt as it goes. While the earliest vacuum robot used random navigation, current models support a more sophisticated mapping system, where the robot can scan the environment with embedded sensors and create a floor plan of the space to guide its cleaning path. Many vacuum robots come with a controller (e.g., bObsweep PetHair+ ) or a smartphone application (e.g., iRobot Roomba 694) to enable users to control the robot and track its status, while some can only be controlled by the physical buttons on the robot (e.g., iRobot Roomba 550)}.
    \item {\textbf{Sidewalk-based delivery robots} are autonomous pedestrian sized robots that deliver items to customers without the intervention of a delivery person. They are usually used for the last-mile delivery after being carried close to the delivery area by a "mothership" van \cite{jennings2019study}. A delivery robot is equipped with cameras for 360-degree vision, ultrasonic sensors, and GPS trackers to autonomously navigate the environment. However, if the robot runs into significant challenges, such as a deep street curb, a human operator can take over the control. A user can order robotic delivery service, track the robot's position, and unlock the cargo compartment via a smartphone application. One example is the Starship Robot, which has a size of \minorchange{26.7" length $\times$} 22.4" width $\times$ 21.8" height. A flagpole extends the robot’s height to 49.1". It can travel at a maximum speed of 3.7 mph \cite{Starship}}.
    \item{\textbf{Drones} are unmanned aircrafts that operate autonomously or are controlled remotely by human operators \cite{wang2016flying}. While originally designed for military use, lightweight drones have been widely adopted for civilian and commercial use, \change{such as taking photos and recording videos}. This type of drones are often in small sizes and equipped with cameras. Users can operate the drones remotely via a controller.}
\end{itemize}

\subsection{Procedure}
Our study consisted of a single session semi-structured interview. Each interview lasted 1 to 1.5 hours. Due to the pandemic, we conducted the study remotely via a Zoom video call. Our interview included three phases, asking about participants' demographic information, experiences with specific robots (i.e., vacuum robots, delivery robots, and drones), and their general experiences and suggestions for robots. We describe each interview phase below. The specific questions are listed in Appendix \ref{sec: interview}. 

\textbf{\textit{Demographics}}. We first asked about participants' demographics (e.g., age, gender, educational level), their visual conditions, their experiences with assistive technologies, and what types of robots they had used or encountered before.  

\textbf{\textit{Experiences with specific robots}}. 
In this phase, we interviewed participants about their experiences with the three mobile service robots. For each type of robots, we asked about participants' experiences and challenges from both direct users' and bystanders' perspectives. 

From the direct users' perspective, we first asked whether the participants had used this type of robots. If yes, we asked about their general use of the robot (e.g., what was the robot brand and model, where, when, and how frequently was the robot used), motivation of using the robot, and their experiences and challenges when interacting with the robot. To better prompt participants to recall and talk about their robot experience, we organized our questions based on robot interaction tasks. Specifically, we divided the robot experience into different tasks and asked participants to describe their experiences with each task. For example, when discussing about the vacuum robots, we asked about participants' interactions, challenges, concerns, and suggestions for different tasks, including setting up the robot, starting the robot, tracking the robot status, locating the robot, charging the robot, verifying the floor cleanliness, and so on. Participants also rated their satisfaction with the robot via a 7-point Likert scale score (7 means completely satisfied, 1 means completely unsatisfied). If the participant had not used such a robot, we asked whether they would want to use such a robot in the future and the reason. 

From the bystanders' perspective, we asked participants whether they had encountered such a robot that was not owned by themselves. If yes, we asked the them to describe their experiences when encountering the robot, including when and where they came across the robot, how often they encountered the robot, how they interacted with the robot, and the challenges they faced. 

By the end, participants gave a 7-point Likert scale score to evaluate their perceived safety around the robot, where 7 indicated completely safe and 1 indicated completely unsafe. Participants also discussed about their suggestions for the robot. We repeated the above steps for each type of robots.

\textbf{\textit{General experiences and suggestions}}. In the last phase, we first asked what other types of robots participants used or came across in their life and their experiences with these robots. We then ended our interview by asking about their general impressions and concerns about the mainstream mobile service robots, such as safety and accessibility issues, as well as their suggestions. 

\subsection{Analysis}
We video recorded the study with the Zoom recording feature and transcribed the interview using the Zoom live transcription function. One researcher on the team reviewed the interview videos and the transcripts manually to correct errors caused by automatic transcription. 
We analyzed the transcripts using qualitative analysis \cite{saldana2021coding, braun2006using}. We first developed codes via open coding. Two researchers coded three identical sample transcripts independently and discussed the codes together multiple times to generate an initial codebook. One researcher then coded the rest of the transcripts based on the initial codebook and listed new codes. The codebook was iterated and updated upon the agreement between the two researchers. \change{The final codebook consisted of 164 unique codes, covering participants' experiences with the robots as direct users and bystanders from different aspects, ranging from accessibility (e.g., visual map, stuck robot), to physical safety (e.g., trip over, cut by propellers), privacy (e.g., capture photos, trespass private property), to security (e.g., robot authentication, delivery inspected by others).}

We then derived themes from the codes. \minorchange{Our research has a specific goal to understand PVI's robot experiences from multiple aspects, including different robot types (personal vs. professional, individual owned vs. shared, low-lying vs. flying), challenges (e.g., accessibility, safety, privacy), and stakeholder perspectives (direct user vs. bystander). With these particular aspects of interest, our theme derivation followed the semantic and deductive approaches, where we focused on specific aspects of the data and identified themes based on the explicit meanings \cite{braun2006using}.} 
\change{Specifically, we sorted and grouped all codes into themes and sub-themes based on the different aspects of interest.} After the initial themes were identified, researchers cross-referenced the original data, the codebook, and the themes, to make final adjustments, making sure that all codes fell in the correct themes.

Our analysis resulted in eight themes: General use of service robots, Motivation of using service robots, Accessibility issues of personal robots, Experiences with shared robots, Safety issues around robots, Privacy concerns around robots, External assistance for robot use, and Expected applications of mainstream robots (see Appendix \ref{sec: theme}). For example, the theme of Safety issues around robots included three sub-themes and 33 codes: (1) Safety concerns as direct users (codes: trip over, low-lying robots, high-flying robot, kicked robots, stuck/stop at middle of the room, watchful gaze, defeat purchase goal, hit by drones); (2) Safety concerns as bystanders (codes: lack of space, hazard, overwhelming navigation, break cane, confuse guide dogs, absence of operators, lack of robot knowledge, buzzing noise, short reaction time, falling drone, cut by propellers, threatening, scared, startling, confused, uncertainty, missed by cane); (3) Suggestions for robot safety (codes: notification from mobile app, AI detects people with disabilities, awareness of cane, collision avoidance, safe distance, safe landing, drone operator training, develop standardized knowledge).



\section{Findings}

\subsection{General Use of Service Robots}

Participants reported various experiences with mobile service robots as direct users and bystanders \change{(Table \ref{tab:demo})}. All 17 participants had experiences with vacuum robots (including both using and encountering the robots), 6 participants for delivery robots, and 12 participants for drones. Moreover, two participants (Tim and Thomas) reported using other types of robots, including a robot mop (Tim) and autonomous driving vehicles (Thomas).

\textbf{\textit{\change{Experience as Direct Users.}}} As direct users, all participants owned or used a vacuum robot, including Roomba, Hoover, Eufy Robovac, and bObsweep PetHair+. \change{Most participants had used vacuum robots for more than a year and they used the robots frequently on a daily or weekly basis. Five participants reported using a vacuum robot everyday, seven used it at least once a week, while five only tried a vacuum robot for limited times. Participants mostly used the robots to clean the entire house, apartment, or workspace (e.g., Edward used a vacuum robot to clean his laboratory).} 

Two participants used delivery robot services. \change{Specifically, Ara used the Starship delivery robot service intermittently since two years ago, and John mentioned using delivery robots four times in the past one year.}
Seven participants had experience flying a drone. \change{However, their drone experience was quite limited. Six participants used a drone once and only Samuel used it twice. Three participants (Meg, Samuel, Reena) tried the drone inside their houses, five had flown their drones outdoors, such as in parks (Rita, Ahmed) and outside of their houses (Sara, Samuel, Nina)}

Among the three types of robots, we found that participants were least satisfied with the drones. When asked to rate their satisfaction with each robot (score ranges from 1 to 7, 7 means completely satisfied, 1 means completely unsatisfied), participants gave drones the lowest score ($mean=1.5, SD=0.78$) but felt somewhat satisfied with the vacuum robots and delivery robots (vacuum robot: $mean=4.93, SD=1.57$; delivery robot: $mean=5.5, SD=2.12$). We elaborate the reasons of participants' dissatisfaction in details in later sections.


\textbf{\textit{\change{Experience as Bystanders.}}} As bystanders, some participants encountered robots that were not owned or used by themselves. Three participants (Meg, Nina, and Reena) came across other vacuum robots at friends’ place. Six participants encountered delivery robots on the sidewalks, with three \change{(Ara, Kat, Rose)} running into them frequently. Nine participants came across drones in public spaces, such as parks and college campus. \change{Four participants only encountered drones once, while five participants encountered them multiple times.} 


\subsection{Motivation of Using Service Robots}

\change{
Similar to sighted users \cite{forlizzi2006service, iocchi2012domestic},} most participants (e.g., Ara, Ahmed, Kat) used service robots because they were convenient and time-saving, enabling multitasking. \change{ 
More importantly, the service robots enabled participants to accomplish tasks that were originally difficult or impossible for them. For example, we found that cleaning the floor was more difficult for PVI than many other housework tasks due to the large physical space it involved (e.g., John, Thomas). This made the vacuum robots particularly valuable to the participants. As Thomas said, \textit{``[The vacuum robot] seemed to have cleaned the area nicely. I always take care of the dishes and everything [by myself]. And, You know, all the sheets and towels and all that kind of stuff. But I was anxious about the floors.''} Interestingly, we also found that the vacuum robots were especially helpful for participants who had guide dogs to clean dog hairs (e.g., Mina, Kat, Reena).}  

Moreover, some participants mentioned that robot service was more cost-effective than human service. \textit{``Well, I like that I didn't have to tip the [delivery] robot. It is nice not to have to pay any extra ... because you don't feel sorry for the robot having to go out in the rain to get the food'' (John).}


\subsection{Accessibility Issues of Using Personal Service Robots}
\label{sec: accessibility}

\subsubsection{\textbf{Controlling robot movements}}
\label{sec: controlRobot}

While benefiting from the service robots, participants faced various accessibility challenges when controlling their personal robots, such as vacuum robots and drones. We identified two primary types of control for personal robots---toggling the robot on and off, and fine-grained control of the robot. We report participants’ interaction experiences, challenges, and strategies for each type of control.

\textbf{\textit{Toggling the robot on and off.}} \change{
Directly pressing the physical buttons on a robot was one major way to start and stop a robot. Like other modern electrical appliances \cite{morris2006clearspeech,fusco2015appliance}, participants found the buttons on many service robots challenging to use}. Five participants reported the difficulty of distinguishing the physical buttons on their vacuum robots since the buttons were all flat (like a touchscreen) and blended into the robot surface without providing any tactile feedback. To use these flat buttons, participants (e.g., Ara, Kat, Reena) had to memorize each button's location and its function, which increased their cognitive load. 
\change{Due to the inaccessibility of the basic robot control, some participants (Ara, Rita) gave up adopting a vacuum robot after several trials.} 

\textbf{\textit{Fine-grained control of the robot.}} Besides the basic on and off functions, most robots enabled fine-grained control. For example, the drones required users to control its movement in real time, and some advanced vacuum robots allowed users to assign a specific room to clean. Compared to simply toggling a robot on and off, such fine-grained control was more difficult to our participants.

\textit{\textbf{(1) Target location assignment.}} Some vacuum robots provided a smartphone application to enable users to control the robot remotely. \change{The most challenging feature in such applications was the Room Mapping feature}---the robot scanned the whole space to generate a visual map, with which the user could divide their house into different sections, name the sections, and assign specific sections to the robot to clean. Seven participants mentioned the difficulty of labeling different rooms on the visual map. Tim described his experience, \textit{``That whole marking and labeling of the individual rooms is not accessible. It's a visual map that shows up on the [phone] screen and you have to distinguish one room from another in the map [and] label them as kitchen, living room, whatever. Because that is inherently a visual process working with the map, I've never been able to do it.''}
    
To make this feature more accessible, \change{Edward and Jen suggested building and labeling the map based on the user's or robot's physical location, \textit{``The only way that I can think of is, if you put the [robot] in the room, and then you somehow tell it I will call the room this. So physically placing the robot in the room, and then I'm kind of [speaking to] the robot, for example `this is the kitchen or the bedroom''' (Jen).}}


\textit{\textbf{(2) Real-time movement control.}} Controlling the drone's movement in real-time was another big challenge to the participants. Participants reported two approaches to controlling the drone's direction and speed: joystick and arrow buttons. 
However, neither approach enabled participants to precisely control the drone. \change{Without any real-time information of the drone's position and status}, participants could not understand the correlation between the joystick's movements (or the duration of button pressing) and the drone's actual behaviors. As Rita explained, \textit{``We couldn't tell how far to the left or to the right [the drone moved]. For example, if we wanted to go out 10 degrees to the right, we couldn't measure that with the joystick.''} This echoed the drone control difficulty reported in Gadiraju et al.'s research \cite{gadiraju2021fascinating}. \change{Moreover, some participants were uncertain about the consequence of releasing the control buttons. \textit{``It's difficult to know what happened when you didn't press a button. Was it staying hovering or was it kind of going down'' (Rita).}}

Avoiding obstacles and landing safely were particularly difficult since participants cannot visually detect obstacles and determine the drone's position in relation to the surrounding environment. As a result, Samuel and Reena mentioned that their drones crashed into things many times, \change{and both Rita and Ahmed's drones crashed and broke the first time they flew them.} Participants thus suggested using AI technology or remote human agents to assist with obstacle avoidance. For example, Reena suggested providing automatic audio descriptions about the drone's surroundings via its camera or having an autopilot option for autonomous navigation. Rita also suggested sending the video stream from the drone camera to a remote human service to get audio guidance.

\subsubsection{\textbf{Tracking robot location and status}} 
\label{sec:trackRobot}
Tracking a robot's location and status (e.g., power level, stuck robot) was another key task when using service robots. Participants mainly leveraged audio and visual indicators from the robots to infer their location and status but encountered various problems. We report their challenges and strategies. 

\textbf{\textit{Tracking by sound.}} One major tracking strategy used by all participants was listening to the noise made by the robots, especially when using vacuum robots and drones. 

For vacuum robots, the absence of noise indicated that the robot was either stuck or had completed its cleaning task. If the robot was not present in its docking station, the participants deduced that the robot was stuck. To help users locate the stuck robot, some advanced robots generated a beeping sound to inform the user. Eleven participants mentioned using the smartphone application or the smart speaker to make the robot beep when they wanted to find the robot. However, \textbf{the worst situation was that the robot ran out of power while getting stuck. The robot was thus not able to generate any notification, making it extremely hard for PVI to find it.} Many participants (e.g., Reena, Kat, Samuel) expressed their concerns.  
\textit{``Sometimes it will go dead while it's stuck [when I'm not at home], because every few minutes, the robot gives another notification, and that runs down its battery, so I wish there was a way for me to tell it like ‘I know you're stuck but I can't do anything right now [since I’m not at home]'' (Reena).}
  
To locate an out-of-power vacuum robot, participants had to physically get down to the floor and scan the whole room. As Thomas described, \textit{``Sometimes I didn't know where [the robot] had stopped. I had to get down on the floor with my cane and swipe around until I found it, and it was tucked away under some table when it stopped by itself.''} To ease this process, some participants (Rita, Tim, Rose) memorized the places where the robot usually got stuck. They either searched these areas first, or prevented the robot from getting into these problematic areas in the first place by blocking its way with obstacles. 

Participants (e.g., Kat, Reena) suggested that the robot should save some power instead of continuously alerting people about its ``stuck'' status. The residual power could then be used to generate an audio indicator upon the user's request. Moreover, Tim mentioned that the smartphone application for the robot should store the latest robot position, so that the users could easily locate the robot even when it was out of power. 

Although the noise and audio feedback were helpful for most PVI, participants who were sensitive to sound (John and Reena) could not tolerate the robot's noise. For example, John had a hearing disability, and the vacuum robot’s cleaning noise was painful for him to hear. He thus only started the vacuum robot when he was out of home. However, the COVID-19 pandemic made the situation more difficult since he could not go out much. As John described, \textit{``I use [the vacuum robot] seldom now since I don’t get out of the house in the pandemic. The vacuum is too loud for me to tolerate it being on in the house. I’m actually hyper sensitive to both light and sound, so I cannot tolerate much light or sound. Unfortunately, pretty much all the robotic vacuums I’ve tried are just way too loud for me to coexist with in the house. So if I cannot leave the house, [I cannot use the robot].''}
    
\change{Compared to a slowly moving vacuum robot used indoors, a fast moving drone was much harder to track, especially in an open space. While the flying noise can indicate the general direction of a drone, participants cannot discern its exact position and moving speed. As Rita said, \textit{``The only way we knew where the drone was was because obviously it made a whirring sound. And that really didn't tell us how far it was. It let us know in basically what direction it was, but it was hard to know how fast it was going.''} Moreover, Meg mentioned that the drone was extremely difficult to locate after it crashed since there was no noise at all.} 

\change{Locating a delivery robot could be even harder for a visually impaired user since it did not generate any audio feedback after it arrived at the destination. Ara described her first experience using a delivery robot, \textit{``The building [I live] has several entrances. Where's this [robot] going to meet me? So I ended up like going outside of the building, walking around, waiting for this thing to show up. I'm just like pacing the building, not knowing where it was actually going to show up, that was problematic. The first encounter I had with it, it was like, didn't say anything, it just sort of stood there...I think I have the luxury of having some vision that I can actually tell where the thing is, I think it would have been more of a hindrance if I couldn't see it at all.''} Some participants further expressed \textbf{strong doubts on the ability of a delivery robot to drop a package off at their doorstep as accurately as a delivery person can}. As Edward described, \textit{``When I have a delivery person, they always put it somewhere on my doorstep. My concern is that because the robot [technology] won't even be there. It may be a little bit away from the doorstep, maybe on the sidewalk leading up to my door somewhere. I have to start really looking around. Most people are just like, oh, there's a package over here let me just go. But I won't have that bird's eye view to pull from, so that's my concern there.''} }
    
\textbf{\textit{Tracking by visual information.}} Our four low vision participants were able to locate the robots by observing the visual information on and around the robots.

\textit{\textbf{(1) Color contrast.}} Some low vision participants (John and Ara) could see a robot with high contrast against its environment, for example, a black robot on a white carpet. 
However, the high contrast did not always exist (e.g., Figure \ref{fig:threerobots}a). As Ara described, \textit{``I wouldn't be able to see where [the vacuum robot] was because it's not contrasting with the carpet, because the carpet is dark as well. So, I had to like fumble around and try to find it, hoping I didn't step on it.''}

To make a robot consistently visible, low vision participants suggested the robot appearance itself presenting a high contrast. As Ara suggested, \textit{``If it were like a black [robot], it could have like a white ring on the outside, or vice versa. So at least the thing has some contrast, and I'd be able to see it.''}


\textit{\textbf{(2) Indicator lights.}} \change{Some robots had indicator lights}, which would either blink or light up in different colors to indicate the robot status. Some low vision participants (Ara and Samuel) used these lights to perceive their robots especially when the noise generated by the robot was dampened by other environmental sounds. 
However, some lights were not visible to low vision participants due to the small size and low luminance \change{(Ara, John)}. As John described, \textit{``There's a indicator [light] for how full the bin is, which is totally useless for me because I can't detect it. It's a tiny, tiny, little indicator.''} Moreover, for low vision participants who were color blind,  it was difficult to distinguish different light colors (Ara). 

\subsection{Experiences of Using Shared Robots} 
\label{sec:sharedRobot}
Unlike personal robots (e.g., vacuum robots), delivery robots introduced unique interactions (e.g., authentication, delivery pick-up) and raised security concerns since the robots were shared among multiple users.

John and Ara were the only two participants who used delivery robot services. They reported different authentication experiences with the robots. John authenticated the delivery robot using text messages, \textit{``You had to send a text message to let it know when you were ready.''} On the other hand, Ara was not aware of any authentication required to pick up her delivery from the Starship robot\footnote{In fact, the food compartment in the Starship robot is mechanically locked and can only be opened via the manufacturer's smartphone application \cite{StarshipAuth}.}. \change{\textbf{Participants expressed security concerns that other people may access their delivery especially when the robot was delivering to multiple users.} As Thomas mentioned, \textit{``How would I know [the delivery robot] is a secure thing. Did anybody get into my food? I don't want to have somebody, sending me something that they've tampered with. I don't want my food inspected or tasted by somebody along the way.''}}


    
Moreover, Ara was concerned about picking up the wrong order from the robot. 
She suggested the robot providing confirmation on whether she opened the right food compartment and took the right item, \textit{``Having [the delivery robot] say, please pick up your food in the slot located wherever...and saying, yes this is your order, this is the right place. [The robot should] just have that interaction and confirmation.''}

\subsection{Safety Issues around Robots}
\label{sec: safety}
Participants emphasized their safety concerns around mobile service robots as both direct users and bystanders. Participants rated their perceived safety for each type of robots (the scale ranges from 1 to 7, where 7 means extremely safe and 1 means extremely unsafe). We found that, while all three types of robots posed safety risks, drones were perceived to be most dangerous (perceived safety of vacuum robot: $mean=5.94, SD=0.80$; delivery robot: $mean=5.40, SD=1.49$; drone: $mean=4.44, SD=2.06$). We elaborate participants' safety perception of different robots from direct users' and bystanders' perspectives respectively. 


\subsubsection{\textbf{Safety concerns as direct users.}} Although the vacuum robot was rated as the safest robot, 15 out of the 17 participants considered it as a potential tripping hazard when it was in the middle of the room. Nine participants even expressed fear of tripping over the low-lying vacuum robots, especially when the robot ran out of power and stopped (or got stuck) at some random places. \change{Mina reported tripping over her robot, Jen had almost tripped over it, and Tim and Meg had accidentally kicked the robot a few times. Mina described her tripping experience,
\textit{``I've tripped over [my vacuum robot]. Because I forget, sometimes it doesn't go back into the [charging] port. It'll just stay where it is. Like, it can't [get to] where the charging port is. And it'll be close to it, but not exactly on it. So I'll trip over it because I don't know it's not there.''} \textbf{To protect their safety, some participants had to pay close attention to the vacuum robot, which conflicted with their original goal of using the robot to save time.} \textit{``I had to keep a watchful gaze on it because I wouldn't want to be stepping on it. So, it pretty much left me tending to the vacuum, when you're really supposed to be doing other things, while the vacuum does its thing. So that it kind of defeats the purpose of having a robotic vacuum'' (Ara).}} 


Compared to the vacuum robots, some participants were more concerned about their safety when using drones. Five out of seven participants (e.g., Meg, Rita) expressed a strong fear of being hit by the drone during use. As Meg mentioned, \textit{``It wasn't really loud enough for me to navigate it without crashing easily. There [were] several close calls of [the drone] attacking me as I didn't move [away from the drone] faster. That goes so fast that there was not a lot of time to react.''} Reena felt it particularly dangerous to fly a drone in indoor environments, \textit{``Outside [the house], I felt safe enough, but inside [the house], I didn't feel safe because it kept on crashing into the ceiling.''} However, Sara did not fear for flying a drone 
since she believed that drone mostly flew at elevated heights beyond human reach, \textit{``If [the drone is] coming close to my head, I need to duck, but usually they are way up there, so yeah, I don't have to worry about [getting hit by it].''}


\subsubsection{\textbf{Safety concerns as bystanders.}} Besides direct user experiences, encountering mobile robots as bystanders also raised safety concerns. We focused on participants' bystander experience with delivery robots and drones since participants rarely had bystander experience with the domestic vacuum robots. 

\textbf{\textit{Encountering Delivery Robots.}} Most participants (except for Summer) felt confused and anxious when they ran into a delivery robot on the sidewalk. \textbf{They perceived the robot as an \textit{unfamiliar} moving obstacle, which can potentially trip them.} \change{Different from many moving objects or vehicles that were directly controlled by a user nearby (e.g., a child riding a bike), \textbf{the absence of direct operators brought more uncertainty to the participants}. As Kat said, \textit{``I just know [the delivery robots] are some things that I shouldn't mess around with because nobody's really there. I don't know if they'll stop or not. If I touched it by accident or something. I don't know if there's anybody watching them.''}

Without clear explanation of the robots' behaviors and intents, participants even felt that the robot ``was confused and did not know what to do'' when it ran into pedestrians (Thomas).} For example, Thomas described how he was distressed by a delivery robot on the sidewalk,
\textit{``I encountered a robotic delivery device on the sidewalk in downtown San Francisco, and I was confused by it. And I think it was confused by me. I heard something on the sidewalk between loud traffic noises. And then, my cane detected [the robot] and it was moving. I think it stopped because my cane touched it. I reached out to see what my cane had touched. And there was a box about the size of a shoe stand polishing box, and I don't know what it was because I was by myself. I realized [since] it was a box, it was probably a delivery box... I was anxious about having heard it, having touched it with my cane and having it available to trip me.''} \change{While using a white cane to scan the environment, Thomas emphasized the risk of the cane missing the robot since it was moving, \textit{``I don't want it to suddenly be there. That would surprise me, especially if I didn't detect it with my cane.''}}

\change{Not only blind pedestrians, the moving robot could also be a hazard for low vision people, especially in visually challenging environments. As Ara mentioned, \textit{``If the sun wasn't in my face. I was able to see it coming. One time the sun was in my face and I did smack into one [delivery robot].''}}

Interestingly, instead of worrying about the robot hurting themselves, \textbf{some participants were concerned about the delivery robot breaking their canes \change{(Ara, John) or confusing their guide dogs (Kat, Rose)}}. As Ara described, \textit{``I really didn't have any trepidation about [the robot] hurting me physically or injuring me, but I definitely had concerns that it would definitely break my mobility [cane] and then I would have a bigger problem.''} \change{Moreover, Kat was nervous about how her guide dog would react to a delivery robot, \textit{``I'm somewhere on the sidewalk. And I have a guide dog and a lot of time, [my dog] is treating it like a moving object. Then I don't know if [my dog] is going to crash on [the robot] or not.''} Since the delivery robots were uncommon objects on the street, guide dogs were not trained to deal with them properly. As Rose mentioned, \textit{``I've heard stories about people who are blind and they have guide dogs, and the Amazon robots, they don't know to move over, so you're kind of in a stalemate since your guide dog won't move over and the robot won't move over.''}}

Given these concerns, participants did not feel comfortable sharing the sidewalk with delivery robots since they took over the already limited and challenging navigation space \cite{bennett2021accessibility}. As Ara explained, 
\textit{``The sidewalks are narrow and really don't have enough space to have a robot running around with other pedestrians. And then you have yourself and people who have a guide dog, cane or whatever it may be, trying to navigate the sidewalk plus listening to these ambient sounds around you...there's definitely a lot of hazards that come with [the delivery robot].''}

\textbf{\textit{Encountering Flying Drones.}} Compared to delivery robots, participants expressed more safety concerns towards drones as bystanders \change{since the flying drones could not be detected by their white canes}. Some participants (e.g., Ara, Ahmed, John) felt scared that the drone could fall and hit them when losing signal or power. Since these drones could be flying at a high speed, the reaction time to these incidents would be minimal (Meg).  
\change{Ara reported that she was almost hit by a drone, \textit{``I was walking around on campus. There happened to be drones flying above one of the buildings, taking photographs for some reason. And one of drones actually fell. And it fell right like five feet in front of me, like if I was further in my distance of travel, that thing will hit me...That was my fear like, oh my god one of these is going to like fall on me and definitely hurt me.''}}
This situation could be more dangerous for PVI because they cannot see the drone to determine the necessary actions to avoid the danger.
As John elaborated, \textit{``A blind person is not going to know when they need to put up their hands on their face defensively...if you don't know what it is like, you don't even know if it has any sharp thing or something on it, and you don't know whether it's propellers are moving, its propellers could cut me.''} 

\change{While most participants feared for drones in general, Rose's safety perception of drones was associated with her knowledge of and trust on the drone operators. As she explained, \textit{``[My concerns] would depend on my familiarity with the user. If it's my friend, I trust him to be a good pilot and not like crashing [the drone] to me or anything. But I think with some random person, I might feel worried. I think it depends a lot on like the pilot.''}}

\textbf{\textit{Suggestions on safety enhancement for bystanders.}} \change{We report participants' feedback to improve their safety around robots. First, to enhance their awareness of approaching robots, all participants suggested the robots providing unique audio signals to alert blind bystanders. For example, Nina expressed the needs for a smartphone application for bystanders to notify them of the presence and purpose of the surrounding robots, \textit{``I want to have an app on my phone that would announce [if] there are drones near your location, and what they're trying to do.''} 
} 

Beyond an alert mechanism, participants (e.g., Ahmed, Thomas, Rose) hoped that the robots were intelligent enough to avoid humans. For example, Thomas wanted the robot to automatically identify people with disabilities on sidewalks, \textit{``I want Waymo (Google's self driving car) and other autonomous vehicles [to] create entities to identify mobility impaired people, old people, blind people, people in wheelchairs, and dog handlers [since] we may not completely interpret accurately the location of the [robots], or how it is intended to interact with us.''}

\change{To achieve more agency to protect their own safety, participants also emphasized the importance of learning basic knowledge of the robots. As Reena said, \textit{``I'm not sure what the etiquette is for [robots], like, I don't know if they cross the street or not. If I'm supposed to treat them like a car or not. Do I wait for them? Do I not? If I'm supposed to do something other than what I normally do, then it needs to become standardized common knowledge, or it needs to tell me.''}}

Moreover, John commented on bystander safety from the policy perspective. He suggested enacting a law that restricted drone operators to experts with sufficient training, ensuring that the drone was piloted in a safe manner and would not hurt any bystanders.


\subsection{Privacy Concerns as Bystanders around Robots}
\label{sec:Privacy}
Participants faced privacy risks around robots. Similar to the privacy perceptions of sighted people \cite{butt2021deployment, LINTAN2021101462, vcaic2018service}, some visually impaired participants (e.g., Ahmed, Samuel, Nina) were concerned about the privacy violation caused by robots, especially drones, since they had the ability to trespass their private property and capture unsolicited information. They felt the drone ``invasive and threatening'' (Thomas). Samuel described his anxiety about drones, \textit{``There is an invasive quality, because it probably has a camera... It's almost like these people are watching us with their drones.''}

Moreover, the privacy risk could be more severe to PVI since they cannot visually detect the drone and determine whether they were being captured. John emphasized the privacy inequity faced by PVI, \textit{``It's especially distressing if you're blind, that somebody else is watching you when you can't be watching them.''}

Participants (e.g., Thomas, Rita) expressed their needs for more information about the robots to protect their privacy, including what information was being captured and why the information was captured. As Thomas said, \textit{``I don't want it coming over and inspecting me or hovering above my head...Also, where and why are big, big issues for me. Why is this thing here? What's it looking at? Am I involved? Is this observing me?''}

\subsection{Assistance to Support Interactions with Robots}

To overcome the accessibility and safety issues posed by service robots, participants sought assistance from both technology and human aids. We report the external assistance they used when interacting with robots. 

\textbf{\textit{AI-based technology.}} Participants reported using AI-based technology (e.g., smart speakers, Microsoft Seeing AI \cite{seeing2020app}) to assist with their interactions with service robots. For example, many participants used smart speakers to control at-home devices, including vacuum robots. Some participants (e.g., Edward, Reena, Summer) preferred using a smart speaker to a controller or a smartphone application since it was efficient and supported hands-free interactions. However, the smart speaker only supported basic robot control, such as starting and stopping a vacuum robot (e.g., Jen, Ahmed). Moreover, some participants (e.g., Sara, Ahmed) had difficulty connecting their smart speakers to their robots. As a result, they turned back to the smartphone application or the control panel on the robot. To avoid the trouble of setting up smart speakers for the robots, Edward suggested embedding a smart speaker in the robot itself. Besides smart speakers, Ara used the Microsoft Seeing AI application on her phone to recognize buttons on her vacuum robot.

\textbf{\textit{Help from sighted companions.}} Similar to prior research \cite{gadiraju2021fascinating, bonani2018my}, eight participants (e.g., Thomas, Rita, Kat) asked for help from their friends and family members when interacting with robots. For example, when using vacuum robots, participants asked for help in various tasks, such as starting the robot (Rose and Rita), connecting the robot to the phone application via WiFi or Bluetooth (Sara), identifying buttons on the robot (Kat and Rita), finding the stuck robot (Meg, Edward, and Nina), and checking the cleanliness of the floor after the vacuum robot finished cleaning (Thomas and Edward).

Moreover, when encountering robots on the street, participants sometimes relied on their sighted companions to help identify and avoid the robots. For example, Kat asked the sighted pedestrians nearby to tell her what the moving object was. Participants also felt safer with their sighted companions when encountering a robot. For example, Thomas described his experience when coming across a drone, \textit{``Fortunately, I was with my sighted walking partner, so I didn't get anxious about it, but if I was out here by myself I would be really worried because it sounds like it's close.''} 

\textbf{\textit{Remote agent service for PVI.}} Seven participants used remote agent service via smartphone or smartglass applications (e.g., Be My Eyes \cite{bemyeyes}, Aira \cite{aira}) to conduct visual tasks, such as locating a stuck vacuum robot (Tim, Reena, Edward), \change{reading robot instructions (Rita), and checking the robot performance (e.g., floor cleanliness, Reena)}. As Tim described, \textit{``I use [Be My Eyes] if I go to the usual spots where [the vacuum robot] is not there and I don't feel like crawling around on my hands and knees [to locate the robot].''}

\textbf{\textit{Customer Service.}} Some participants (e.g., Tim, Ara, Reena) mentioned that the robot instructions were in the form of printed manual books, which were inaccessible. Hence, to better use their vacuum robots, some participants (Tim, Rita) sought help from the customer service. \change{However, \textbf{the sighted agents provided only visual cues (e.g., colors of the robot components) as references during the communication, which were not helpful and frustrated the participants.} As Rita said, \textit{``So you spend the first 30 minutes trying to say `no, I don't see what you're talking about,' but they keep asking, `okay now do you see it,' and you keep saying the same thing over and over again.''}}

    

Instead of using visual references, Tim emphasized the importance of clarifying the orientation of the robot and describing a robot component using tactile landmarks, \textit{``[The agents] are sighted and I'm blind and the landmarks that they are using and trying to describe aren’t useful to me. For example, when you're going to replace the brushes underneath the [vacuum] robot, there's this tiny little lever that you have to squeeze to be able to open [the robot] and get the brushes. I couldn't get them to describe that tiny little lever... they just kept saying, Oh, it's the blue lever... That doesn't do me any good. I need you to tell me, is [the lever] near the left wheel or the right wheel, and depending on which way the robot is facing, you know, which way is left and which way is right.''}  

Moreover, Tim expressed his disappointment that the customer service would not take his suggestions on improving the accessibility of the robot because of copyright infringement, \textit{``There's a large, large component of the technology built into this that I can't use. And the fact that, you know, when I call them and say hey, here's the problem that I have and here's, you know, as a blind person, a couple of easy solutions that I think would fix it, and they refuse to listen to me, claiming that it could be a patent infringement if I give them an idea, and they run with it and build it into the robot, they're claiming I could turn around and sue them for copyright infringement.''}


\section{Discussion}

Our research presents the first thorough exploration of PVI's experiences with different types of mobile service robots. Unlike most prior work that focused on the direct use of the robots \cite{gadiraju2021fascinating, forlizzi2006service}, our study paints a more comprehensive picture by taking into account both the direct users' and bystanders' perspectives. \change{We answer the two research questions presented in the Introduction.  

From the direct user's perspective, our study thoroughly revealed PVI's challenges, strategies, and needs when using and controlling different service robots (RQ1). Some of our findings confirmed the insights from prior research, such as the inaccessibility of the control panels on electrical appliances \cite{fusco2015appliance, morris2006clearspeech} and the difficulty of controlling drone movements accurately in real-time \cite{gadiraju2021fascinating}. Beyond that, we uncovered \textit{more detailed evidence} of PVI's challenges and coping strategies during their interactions with robots. For example, besides the basic button control issues, the advanced room mapping feature for the vacuum robot further marginalized PVI since the generated map was too visual (Section \ref{sec: controlRobot}); while PVI were able to track the position of a vacuum robot by its noise, the robot can easily get stuck and out-of-power, thus becoming extremely difficult to locate and posing tripping risks (Section \ref{sec:trackRobot}). In addition to \textit{personal robots} used by individuals, our research also explored PVI's experiences with \textit{shared robots} (e.g., delivery robots), highlighting their security concerns (e.g., delivery being accessed and inspected by other people) and issues with finding the robot and picking up the right order (Section \ref{sec:sharedRobot}).

From the bystander's perspective, our study identified unique safety and privacy challenges faced by PVI (RQ2). We found that mobile service robots were a source of fear and anxiety for visually impaired bystanders due to the lack of accessible feedback and standardized knowledge of the robots. Unlike most other moving objects controlled by a user nearby (e.g., a tricycle ridden by a child), the absence of immediate operators of autonomous robots brought more uncertainty and concerned the PVI. Moreover, beyond their own safety, PVI were worried about the robots breaking their canes or confusing their guide dogs (Section \ref{sec: safety}). 
Finally, our research revealed the privacy inequity faced by visually impaired bystanders in front of robots that incorporated various sensors (e.g., camera, Section \ref{sec:Privacy}).

}


Based on participants' experiences and suggestions, we derive design considerations to inform more accessible and safe mobile service robots for PVI.

\subsection{Accessible Interactions with Robots as Direct Users }  
Inaccessible feedback and control were the major reasons that hindered effective interactions between PVI and robots. Based on our findings, we discuss four directions to enhance PVI's interactions with service robots as direct users.

\textbf{\textit{Incorporating alternative and customizable feedback.}} \change{Similar to many electrical appliances in people's daily life (e.g., microwave),} current robot interfaces mainly rely on visual feedback to communicate information to users. \change{Hence, insights from prior research on providing audio and haptic feedback alternatives on appliance interfaces \cite{fusco2015appliance, guo2017facade} should also be applied to the mainstream robots to foster accessibility for users with diverse abilities.} Additionally, our study showed that different users had different preferences on feedback modalities. For example, while audio feedback was commonly used by most participants to infer the robot's status, it frustrated people who were highly sensitive to sound (e.g., John). As such, haptic feedback (e.g., vibrations, textures) was preferred in this situation. With each type of feedback having its pros and cons, robot designers should consider providing users more flexibility to customize the feedback based on their abilities and preferences. 

\textbf{\textit{Enhancing visual design for low vision.}} \change{Unlike blind people, we confirmed that low vision people can leverage visual feedback to perceive the surroundings \cite{zhao2015foresee, zhao2018looks, szpiro2016finding}.} However, most visual cues on robots are not visible enough to low vision users due to their small size and low contrast, preventing low vision people from leveraging their residual vision. Besides adding more feedback alternatives, robot designers should consider making the visual design of the robots more accessible to low vision users. For example, the robot appearance itself should reflect high contrast to eliminate the impact of the background environment on the robot's visibility. The indicator lights should be bigger and brighter, and adopt different blinking patterns instead of colors to indicate different robot status, thus diminishing the barriers for people who are color blind.  

\textbf{\textit{Enabling fine-grained control.}} Fine-grained control of robots is an important task for human-robot interaction (HRI). \change{However, the limited audio feedback (e.g., the robot noise) generated by current robots was not informative and precise enough for PVI to accurately control their robots.} Enabling fine-grained control for PVI is challenging since it requires both accurate localization of the robot and real-time feedback to communicate such information. 
In fact, prior research in HRI has investigated computer vision algorithms for robot tracking and collision detection \cite{patel2021event, schmidt2021real}. Gadiraju et al. \cite{gadiraju2021fascinating} has also explored PVI's preferences on the input and output modalities when piloting a drone, however, no system has been implemented and evaluated. Future research should explore how to combine the state-of-the-art recognition technology with accessible feedback to enable precise robot control for PVI. 

\textbf{\textit{Providing accessible instructions.}} Besides refining the robot design itself, how to provide accessible instructions to reduce PVI's learning curve is another important aspect that needs attention from the designers and manufacturers. Our study uncovered two major issues with the current robot instructions. First, the instructions were presented in an inaccessible format that was difficult for PVI to perceive. Second, the instruction contents were visually oriented, providing only visual cues to describe the robot components, which were useless for PVI. To enable PVI to fully benefit from the robot, it is important for manufacturers to provide accessible instructions. For example, robot instructions should provide alternative audio or Braille versions, and customer service agents should be trained to properly communicate with people with disabilities and provide suitable guidance.

\subsection{Space Sharing with Robots as Bystanders}
\label{dis: sharedRobots}
\change{As bystanders around robots, PVI are mostly concerned about their safety and privacy. In this section, we discuss design implications to protect visually impaired bystanders from robots.}

\textbf{\textit{Conveying robot intents.}} Effectively communicating the robot intents to bystanders is an important topic in the HRI field to ensure bystanders' safety. HRI researchers have designed different methods to visualize the robot's moving trajectories for sighted bystanders \cite{watanabe2015communicating, walker2018communicating}. However, our study showed that visually impaired bystanders need more information to feel safe and comfortable around a mobile robot. Based on our findings, we summarize the information required by PVI: (1) Existence of the robot: whether there is a robot nearby; (2) Proximity: how far away the robot is; (3) Movement status: whether the robot is moving or when it will start moving; (4) Moving information: in which direction the robot is moving to or intends to move, and at what speed the robot is moving. Unlike prior work that focused on visualizing the robot's intents, robot designers should explore how to effectively convey this information to PVI in an accessible and unobtrusive manner, \change{for example, using earcons \cite{brewster1993evaluation} to present the existence of different robots.}

\textbf{\textit{Communicating privacy information.}} Bystanders face privacy risks since they lack the knowledge of the robot's design and capability. \change{The risks can become more severe for visually impaired bystanders since they cannot visually detect the sensors on the robots (e.g., cameras, microphones), which hinder them from estimating and avoiding the potential risks. We summarize three key types of information that can help PVI better perceive the robot and protect their privacy}: (1) Existence of sensors: whether and what sensors are present on the robot; (2) Data collection scope: what range of data the sensor is capturing (e.g., the field of view of the camera), and whether the visually impaired bystander is involved in the data collection; (3) Purpose of data collection: why the robot is being used and why specific information is collected. While prior work in privacy has explored sighted bystanders' privacy perceptions and expectations around robots, such as drones \cite{wang2016flying, yao2017privacy}, no research has deeply investigated the unique privacy perceptions and needs of bystanders with visual impairments. Privacy enhancement mechanisms are also needed to accessibly convey the sensor information on robots to foster privacy equity for bystanders with visual impairments. 

\textbf{\textit{Policies to regulate robot use.}} Laws and policies are also needed to ensure the proper use of mobile robots in public spaces. Both our study and prior work \cite{bennett2021accessibility} have shown that mobile service robots occupy public spaces, creating navigation difficulties for people with disabilities. Some states in U.S. have passed laws that limit robot use in public spaces. For example, San Francisco Public Works Code, Article 15, Section 794 \cite{LawDelRobotSF}, rigorously restricts the use of delivery robots in public, including limiting the number of delivery robots to nine and the highest speed to three mph at any given time in the entire city. This law also requires companies to receive permits to operate delivery robots in public with the presence of trained human operators at all times. Contrarily, some states have granted mobile robots more rights to support the development of this technology. For example, Pennsylvania's Title 75 (Vehicle Code) \cite{LawDelRobotPA} gives delivery robots equal rights as ``pedestrians.'' Such less restrictive laws have created uproar amongst pedestrians who complain about the potential safety threats posed by these robots \cite{LawPA}. With the growth of the service robot market, the government should enforce responsible robot deployment and use in public via regulations to protect the right of all bystanders, including bystanders with disabilities. \change{For example, autonomous robots should be tested by people with disabilities before deployment in public, and sufficient outreach should be required to educate people about basic robot knowledge before public deployment.}  


\subsection{Ensuring Security for Shared Robots} 
Compared to personal robots used by a single user, robots shared by multiple users (e.g., delivery robots) pose unique security concerns to PVI. 
To enable accessible authentication for PVI, prior research has explored the accessibility issues and security concerns faced by PVI when authenticating to websites and applications on computers and smartphones \cite{dosono2015m, hayes2017they}. Accessible authentication mechanisms, such as UniPass \cite{barbosa2016unipass}, have also been developed to help PVI securely and easily log into their computers and smartphones. However, it is unclear whether such authentication methods could be applied to robots directly. To ensure a secure robot sharing experience, future research should explore PVI's authentication needs in the context of shared robots, determining whether \change{and what} new methods would be needed for accessible and secure robot authentication. 
    
    
\subsection{Context-aware Robots}

\change{The advance of AI technology presents new opportunities for robot accessibility. In our study, PVI hoped the service robots to be aware of the users' context, including their abilities, situations, and needs, to provide more intelligent assistance.} For example, a delivery robot should recognize their disability, thus automatically avoiding them on the street. In addition to disability awareness, other contextual information about the users and environments could also be helpful. For example, \change{a stuck vacuum robot should consider the user's presence to decide the timing to emit alerts.} 
However, collecting users' contextual information can cause privacy violation, and the problem could be more severe to PVI. It is thus important for designers and developers to consider the trade-off between user privacy and data collection for intelligent service. One potential solution is to ensure data collection transparency by accessibly communicating all data collection and recognition information to PVI (as discussed in Section \ref{dis: sharedRobots}).




\subsection{Limitations and Future Work}
Our study has limitations. First, we focused on PVI's experiences with three types of mobile service robots (i.e., vacuum robots, delivery robots, and drones) to derive general robot design considerations. While our robot selection considered both popularity and coverage (e.g., personal vs. shared robots, low-lying vs. flying robots), there could be specific robot types that we have not covered. In the future, we will expand our study to more diverse robot types, such as social robots and humanoid robots, to construct a more comprehensive understanding of PVI's robot experiences. Second, the number of participants for each type of robots were not equally distributed in our study: all participants had experience with vacuum robots \change{as direct users (few had bystander experience)}, while only partial participants used or encountered drones and delivery robots, which resulted in relatively limited data for these robots. This is not surprising given that the vacuum robot technology is more mature and widely used than drones and delivery robots. As service robots become increasingly popular, we plan to recruit more participants for these emerging robots (e.g., delivery robots). Moreover, we conducted a remote interview study because of the COVID-19 restrictions. In the future, a contextual inquiry study that observes PVI's interactions with service robots will be considered to enrich our data.
\section{Conclusion}
In this paper, we thoroughly explored PVI's experiences with the emerging mobile service robots from both direct users' and bystanders' perspectives. Our study revealed various accessibility, safety, and privacy issues faced by PVI, as well as the strategies and assistance they adopted to interact with the robots. Design considerations have been distilled from multiple aspects, such as accessible control of personal robots, safety around public robots, and security for shared robots. Our study sheds light on the design of more accessible and safe robots for people with visual impairments. 

\begin{acks} 
We thank the National Federation of the Blind for helping us recruit for our study, as well as the anonymous participants who provided their perspective.
\end{acks}

\bibliographystyle{ACM-Reference-Format}
\bibliography{sample-base}

\appendix
\renewcommand\thefigure{\thesection.\arabic{figure}}    
\setcounter{figure}{0}  
\renewcommand\thetable{\thesection.\arabic{table}}    
\setcounter{table}{0}  
\section{Appendix}

\subsection{Pre-Screening Questionnaire}
\label{appendix: screen}
Respondents were eligible to participate in our study if they were 18 years old or above, had visual impairments, and had experiences with at least one of the three types of robots (i.e., vacuum robots, delivery robots, drones).
\begin{enumerate}
    \item How old are you?
    \item Are you blind or low vision?
    \item Could you please describe your visual ability?
    \item Do you have any experience with robots? Vacuum robots? Delivery robots? Drones? Other robots? 
\end{enumerate}

\subsection{Interview Questions}
\label{sec: interview}

\subsubsection{Demographic Questions}

\begin{enumerate}
    \item Are you legally blind?
    \item What is your visual impairment diagnosis?
    \item Do you have functional vision?
   
    If Yes:
    \begin{enumerate}
        \item What is your visual acuity?
        \item What is your field of view?
        \item Do you have light perception?
        \item How is your color vision?
    \end{enumerate}
    \item What is your education level?
    \item Do you live alone or with your family?
    \item Do you live in an urban area or a rural area?
    \item What assistive technology do you use in daily life?
\end{enumerate}

\subsubsection{Introductory Question}
Participants were given a brief introduction of the three types of robots (i.e., vacuum robots, delivery robots, and drones) and then were asked the following question:

What robots have you used or come across before?

Based on participants' answer to this question, we asked them follow-up questions about their experiences with each specific type of robot below. 

\subsubsection{Questions about vacuum robots}

\begin{enumerate}
    \item \textbf{Experiences as direct users:} Have you ever owned or used a vacuum robot?
    
    If No:
     \begin{enumerate}
        \item Would you like to own/use one in future? Why?
    \end{enumerate}
    
    If Yes:
    
    \begin{enumerate}
        \item Could you please tell us the name (brand and model, if possible) of the robot?
        \item When did you get this robot?
        \item What do you use the robot for?
        \item Where do you use the robot?
        \item What can your robot do? What features does your vacuum robot support? 
        \item How frequently do you use the robot?
        
        \item Task-driven questions: 
        
        \begin{itemize}
            \item Robot Initiation:
                \begin{itemize}
                    \item How do you initiate/set-up the robot?
                    \item What challenges do you face while performing this task?
                    \item What strategies have you used to overcome the challenges? 
                    \item How do you want to improve the robot to better support robot initiation?
                \end{itemize}
                
            \item Robot Status Tracking:
                \begin{itemize}
                    \item How often does the robot get stuck?
                    \item Do you have any difficulties figuring out when the robot gets stuck? What are the difficulties?
                    \item If the robot gets stuck or has errors, how do you deal with it?
                    \item If the power gets exhausted in between the task, what do you do? What does the robot do?
                    \item Did you face any other challenges while tracking the status of the robot apart from the ones discussed so far?
                    \item What strategies have you used to track the status of the robot?
                    \item How do you want to improve the current robot design to better understand the status of the robot? 
                \end{itemize}
                
            \item Robot Location Tracking:
                \begin{itemize}
                    \item Do you face any difficulties to understand and track where the robot is? What are the difficulties? 
                    \item What strategies have you used to track the location of the robot?
                    \item How do you want to improve the robot design to better track the location of the robot?
                \end{itemize}
                
            \item Robot Termination:
                \begin{itemize}
                    \item How do you stop the robot?
                    \item Have you faced any difficulties in this task?
                    \item What strategies did you use to overcome the challenges?
                    \item How do you want to improve the robot design to better support this task?
                \end{itemize}
            
            \item Cleanliness Verification:
                \begin{itemize}
                    \item How do you evaluate the performance of the robot?
                    \item What challenges do you face while performing this task?
                    \item What strategies do you use to overcome the challenges?
                    \item How do you want to improve the robot design to better support this task?
                \end{itemize}
                
            \item Robot Maintenance:
                \begin{itemize}
                    \item How do you know when to charge the robot?
                    \item How do you charge the robot?
                    \item What’s your experience with the maintenance of the robot (e.g., changing the filter, etc.)?
                    \item How do you want to improve your experiences with the maintenance of the robot?
                \end{itemize}
    \end{itemize}

         \item How would you rate your satisfaction of using the robot? (7-point Likert scale, where 7 means completely satisfied and 1 means completely unsatisfied.)
            
    \end{enumerate}  
    
        \item \textbf{Experiences as bystanders:} Have you encountered with a vacuum robot that is not used by you? 
        
        If Yes:
            \begin{enumerate}
                \item How frequently do you encounter such experiences?
                \item When do you usually encounter the robots?
                \item Where do you usually came across the robot(s)?
                \item Could you describe your experience when encountering the robot? (Focus on discussing the most unforgettable or most recent experience.)
                \begin{itemize}
                    \item What was the robot doing?
                    \item What were you doing at that time?
                    \item What difficulties or concerns did you have around the robot? 
                    \item How do you want to improve your experiences with the robot as a bystander?
                \end{itemize}

            \end{enumerate}
            
        \item \textbf{General experiences}    
        \begin{enumerate}
            \item Did the robot cause any physical harm to you or anyone you know (tripping injuries, etc.)? Could you describe the situation?
            
            \item How would you rate your safety around the robot? Why do you feel so?  (7-point Likert scale, where 7 means completely safe and 1 means completely unsafe.)
            
            \item How would you want the robot to be improved to better guarantee your safety?
            
            \item In general, What do you like about the robot? Why?
            \item In general, what do you not like about it? Why?
            \item How do you want to improve the robot to better serve your needs?

        \end{enumerate}
\end{enumerate}

\subsubsection{Questions about delivery robots and drones}
The questions for delivery robots and drones are similar to the vacuum robot questions. The only difference is that we focus on different tasks. The tasks for delivery robots include: (1) Robot status and location tracking; (2) Robot greeting; (3) Robot authentication; (4) Order pick-up. 
The tasks for drones include: (1) Robot Initiation; (2) Robot control; (3) Robot location tracking; (4) Robot status tracking; (5) Robot maintenance.

\subsubsection{Exit Interview Questions}
\begin{enumerate}
    \item Have you used any other types of robots? If yes, what robots have you used? Please describe your experiences.
    \item We have gone through three types of mainstream robots---vacuum robots, delivery robots, and drones. What are your general suggestions that the robot designers should consider to make the robots more: (a) safe? (b) accessible?
       
\end{enumerate}

\subsection{Theme Table}
\label{sec: theme}
\small
\begin{table}[h]
  \caption{Summary of Themes and Sub-themes}
  \label{tab:theme}
  \begin{tabular}{l l}
    \toprule
    Themes & Sub-themes\\
    \midrule
    General Use of Service Robots & 
    Robot type, brand, and model \\
      & General robot experience as direct users \\
      & General robot experience as bystanders\\
    \hline
    Motivation of Using Service Robots & Support daily tasks \\
     & Be cost-effective \\
     & Overcome disability-related constraints \\
    \hline
    Accessibility Issues of Personal Robots & Toggling robots on and off\\
     & Fine-grained control of robots \\
    & Tracking robot location \\
    & Tracking robot status \\
    \hline
    Experience with Shared Robots & Robot authentication and order security\\
    & Robot sanitation \\
    & Order pick-up\\
    \hline
    Safety Issues around Robots 
    & Safety concerns as direct users \\
     & Safety concerns as bystanders \\
   & Suggestions for robot safety \\
    \hline
    Privacy Concerns around Robots & General privacy concerns around robots\\
      & PVI's unique privacy concerns \\
      & Information needs for privacy enhancement \\
    \hline
    External Assistance for Robot Use & AI-Based technology\\
      & Help from sighted companions\\
      & Remote agent service for PVI \\
      & Customer Service \\
    \hline
    Expected Applications of Mainstream Robots & N/A\\
  \bottomrule
\end{tabular}
\end{table}

\received{January 2022}
\received[revised]{April 2022}
\received[accepted]{August 2022}

\end{document}